\documentstyle[prc,aps,floats,epsf]{revtex}

\def\thalf{{\textstyle{\frac{1}{2}}}}
\def\thrhalf{{\textstyle{\frac{3}{2}}}}

\def\ttwoth{{\textstyle{\frac{2}{3}}}}
\def\tfourth{{\textstyle{\frac{4}{3}}}}
\def\tquar{{\textstyle{\frac{1}{4}}}}

\newcommand{\ord}[1]{${\cal O}(Q^{#1})$}
\def\slat{\mbox{$\scriptstyle /$}}
\begin{document}

% titlepage stuff
%
% uncomment \draft to have PACS numbers appear
\draft

% put preprint number(s).
\preprint{\vbox{\hfill NUC-MINN-02/6-T}}
%                \null\hfill OSU--96--612}}

\title{Low Energy Pion--Nucleon Scattering in the Heavy Baryon and 
Infrared Schemes}

\author{Kenichi Torikoshi and Paul J. Ellis}
\address{School of Physics and Astronomy,
     University of Minnesota, Minneapolis, MN\ 55455.}

%\date{DRAFT: \today}
\maketitle

\begin{abstract}
We study pion--nucleon scattering  with a chiral Lagrangian
of pions, nucleons, and $\Delta$-isobars to order $Q^3$,
where $Q$ is a generic small momentum. We compare the results
from heavy baryon chiral perturbation theory with those from the 
infrared regularization scheme. While the former provides a reasonable 
fit to the data out to a pion c.m. kinetic energy of 100 MeV, the latter is
only able to fit up to 40 MeV and even then the parameters obtained are 
unreasonable. Difficulties with the infrared scheme in the 
$u$-channel are discussed.
\end{abstract}

\vspace{20pt}

%\narrowtext

\section{Introduction}

Pion-nucleon scattering is a fundamental process that one would like to 
describe using the low-energy realization of quantum chromodynamics, namely
chiral perturbation theory \cite{WEINBERG79,GASSER84}. This is an attractive 
approach because it not only embodies chiral symmetry, which is 
fundamental to low-energy physics, but also offers a systematic expansion 
in powers of the momentum. Further it  ensures unitarity order by order. 
Gasser and Leutwyler \cite{GASSER84} have shown that chiral perturbation
theory works nicely for mesons and high accuracy can now be achieved for 
the $\pi\pi$ scattering lengths \cite{pipi}.

However, the power counting fails when baryons are
introduced \cite{GASSER88}. The power counting can be restored in 
heavy baryon chiral perturbation theory (henceforth referred to as HB)
\cite{JENKINS91} where the heavy components of the  baryon fields 
are integrated out. An alternative -- the infrared regularization scheme 
(henceforth referred to as IR) -- has been proposed \cite{bl}, based on 
the ideas in Ref. \cite{et}. This preserves the chiral power 
counting and has the advantage that it is manifestly Lorentz invariant and 
avoids the voluminous effective Lagrangian of the HB approach.  
Generally favorable results have been obtained with the IR scheme in a 
number of applications \cite{irapp}. By suitable approximation of
the IR expressions the HB formulae can be regained. A recent review of the 
HB and IR formalisms has been given by Mei\ss ner \cite{mrev}.

The first fit to the pion-nucleon scattering phase shift data using chiral 
perturbation theory was carried out in Ref. \cite{et}, henceforth referred 
to as I, using the HB scheme.
In addition to the nucleon and pion fields, the $\Delta$ resonance field 
was included explicitly since the intent was to fit out to energies in the 
$\Delta$ resonance region. The calculation was carried
to ${\cal O}(Q^3)$, where $Q$ is a generic small momentum scale,
and it was found possible to obtain a reasonable fit up to energies slightly 
below the $\Delta$ resonance. Subsequently Fettes and Mei\ss ner carried out 
several HB studies both with \cite{fm3nd} and without \cite{fm3n,fm4} explicit 
inclusion of the $\Delta$ field. The calculation with the $\Delta$ field
to ${\cal O}(Q^3)$ appeared to be a little better than the 
${\cal O}(Q^4)$ calculation without it; for further discussion see 
Ref. \cite{erev}. Fettes and Mei\ss ner \cite{fmch} have also studied 
isospin violation in the $\pi-N$ system, although here we shall focus on the 
isospin symmetric case. All of these calculations were carried out 
with the HB approach and it is natural to examine the IR method in this 
context. Becher and Leutwyler \cite{blpi} have used the IR approach
in the sub-threshold region, where the HB scheme is inappropriate, 
and found that the 
IR representation of the scattering amplitude was not sufficiently accurate 
to allow the extrapolation of the experimental data to this region.
The purpose of the present work is to study the IR scheme in the physical 
region at \ord{3} and compare it with the HB approach in order to see 
whether an improved fit to the phase shift data can be obtained.

The organization of this paper is as follows. In Sec. II we review
our notation for the effective Lagrangian and discuss the \ord{3} 
calculation of the scattering amplitudes in the IR scheme and their 
reduction to HB form. Formulae for the $\sigma$ term and effective 
vertex couplings are also given.
In Sec. III our fit to the phase shift data is described, first for the 
HB scheme and then for the IR approach. Our conclusions are presented in 
Sec. IV. Expressions for the IR integrals mentioned in the text are given 
in the Appendix, together with their HB reductions.

\section{Formalism}
\subsection{Effective Lagrangian}

In the IR or HB schemes the Feynman diagrams for $\pi N$ scattering 
follow Weinberg's power counting rule \cite{WEINBERG90}. Define
\begin{equation}
      \nu = 1+2 L 
          + \sum_{i} V_i \Big (d_i + \case{1}{2} n_i -2\Big) 
                \ , \label{eq:chcnt}
\end{equation}
where  $L$ is the number of loops, $V_i$ is the number of
vertices of type $i$ characterized by $n_i$ baryon
fields and $d_i$ pion derivatives or pion mass factors. 
Then a given diagram is of leading order $Q^\nu$, where $Q$ is a small 
or ``soft" momentum scale, for example, the pion mass $m$, the pion 
momentum or the mass splitting between the $\Delta$-isobar and the 
nucleon, $\delta=M_\Delta-M$.

We briefly recapitulate the effective chiral Lagrangian given in I.
Characterising it by $d_i + \case{1}{2} n_i$, the order $Q^2$ part is 
\begin{eqnarray}
  {\cal L}_2 &=& \bar{N} ( i\rlap/{\mkern-3mu {\cal D}}
               +g_{\rm A}\gamma^\mu \gamma_5 a_\mu - M ) N 
           +\tquar f_\pi^2  {\rm tr}\, (\partial_\mu U^\dagger
                         \partial^\mu U)
           +\tquar m^2f_\pi^2 {\rm tr}\,(U + U^\dagger -2)
       \nonumber  \\
      & & 
          +  \bar{\Delta}_\mu^a
           \Lambda^{\mu\nu}_{ab} \Delta_\nu^b
       + h_{\rm A} \Big ( \bar{\bbox{\Delta}}_\mu
         \bbox{\cdot a}^\mu  N
       + \bar{N}
           \bbox{a}^\mu\bbox{\cdot \Delta}_\mu \Big)
         + \tilde{h}_{\rm A}
             \bar{\Delta}_\mu^{\, a}
                  \gamma^\nu \gamma_5 a_\nu
                    \Delta^\mu_a   \ ,   \label{eq:L2}
\end{eqnarray}
where the trace is taken over the isospin matrices.
The isotriplet pion fields enter in the $SU(2)$ matrix
\begin{equation}
 U(x) \equiv \xi ^2 =   \exp (2i\pi(x)/ f_{\pi})
                              \ , \label{eq:U-def}
\end{equation}
where $f_{\pi}$
is the pion decay constant and 
$\pi(x) \equiv \bbox{\pi}(x)\, \bbox{\cdot}\,{1\over 2}\bbox{\tau}$,
with $\bbox{\tau}$  denoting the Pauli matrices.
The axial vector field
$a_{\mu}(x)$ and vector field $v_{\mu}(x)$ are defined by
\begin{eqnarray}
a_{\mu}     & \equiv & -{i \over 2}(\xi^{\dagger} \partial_{\mu} \xi -
    \xi \partial_{\mu}\xi^{\dagger} )
          = a_{\mu}^\dagger= \thalf\bbox{a}_\mu\bbox{\cdot\tau}
     ={1\over f_\pi}\partial_{\mu}\pi
             - {1\over 3f_\pi^3}  \pi [\pi,\partial_{\mu}\pi] +\cdots    
          \ ,\label{eq:adef}  \\[4pt]
v_{\mu}     & \equiv & 
           -{i \over 2}(\xi^{\dagger} \partial_{\mu} \xi +
    \xi \partial_{\mu}\xi^{\dagger} )
         = v_{\mu} ^\dagger   = \thalf\bbox{v}_\mu\bbox{\cdot\tau}
         =-{i\over 2f_\pi^2}
           \left(1-{\pi^2\over 3f_\pi^2}\right) [\pi,\partial_{\mu}\pi]+ \cdots 
            \ , \label{eq:vdef}
\end{eqnarray}
both of which contain one derivative. The covariant derivative on the 
nucleon field ${\cal D}_\mu N = \partial_\mu N + i v_\mu N$.
For the $\Delta$ resonance we have introduced  an isovector field
$\bbox{\Delta}_\mu = \bbox{T} \Delta_\mu$
in terms of the standard $2\times 4$
isospin $\thrhalf$ to $\thalf$  transition matrix and the labels $a$ and 
$b$ in Eq. (\ref{eq:L2}) are isospin indices. The
kernel tensor in the $\Delta$ kinetic energy term is
\begin{equation}
    \Lambda^{\mu\nu} = -(i \rlap/{\mkern-3mu {\cal D}} 
                - M_\Delta)g^{\mu\nu} 
              +i(\gamma^\mu {\cal D}^\nu +\gamma^\nu {\cal D}^\mu)
              -\gamma^\mu(i\rlap/{\mkern-3mu {\cal D}}+M_\Delta)\gamma^\nu
                        \ ,
\end{equation}
suppressing isospin indices, and the covariant derivative is defined by
\begin{equation}
{\cal D}_\mu \bbox{\Delta}_\nu 
            = \partial_\mu\bbox{\Delta}_\nu
                     + i  v_\mu  \bbox{\Delta}_\nu
                   -   \bbox{v}_\mu \times \bbox{\Delta}_\nu\ .
\end{equation}
We have chosen the simplest form for the $\pi N\Delta$ interaction 
in Eq. (\ref{eq:L2}) since modifications of the type suggested by 
Pascalutsa \cite{pasc} or modifications of the standard off-shell 
$Z$ parameter can be absorbed in the other parameters of the 
Lagrangian \cite{TE96}.

The order $Q^3$ and $Q^4$ parts of the Lagrangian are:
\begin{eqnarray}
  {\cal L}_3 &=&  { \beta_\pi \over M} \bar{N} N 
         {\rm tr}\, (\partial_\mu U^\dagger \partial^\mu U)
          -{\kappa_\pi\over M} \bar{N}
               v_{\mu\nu}\sigma^{\mu\nu} N
                     \nonumber  \\
      & & +{\kappa_1\over 2 M^2} i\bar{N}
               \gamma_\mu 
               \stackrel{\leftrightarrow}{\cal D}_\nu N
             {\rm tr}\, (a^\mu a^\nu)
          +{\kappa_2\over M} m^2 \bar{N} N\, 
            {\rm tr}\,(U + U^\dagger -2) +\cdots
                   \ ,     \label{eq:L3}  \\[4pt]
  {\cal L}_4 &=& 
             {\lambda_1\over M} m^2 \bar{N} \gamma_5
                (U - U^\dagger ) N
             + {\lambda_2\over M^2} \bar{N}
                        \gamma^\mu D^\nu v_{\mu\nu} N
                     \nonumber  \\
      & & 
           +{\lambda_3\over M^2} m^2 \bar{N} \gamma_\mu
               [a^\mu, \, U - U^\dagger] N
          +{\lambda_4\over 2 M^3} i\bar{N}
               \sigma_{\rho \mu} 
               \stackrel{\leftrightarrow}{\cal D}_\nu N
             {\rm tr}\, (a^\rho D^\mu a^\nu)
                     \nonumber  \\
      & &
          +{\lambda_5\over 16M^4} i\bar{N}
         \gamma_\rho 
            \{\stackrel{\leftrightarrow}{\cal D}_\mu,
            \stackrel{\leftrightarrow}{\cal D}_\nu\} \tau^a N 
          \,{\rm tr}\,(\tau^a[D^{\rho}a^\mu,a^\nu])
              \nonumber  \\
      & &
         +{\lambda_6\over 2M^2} m^2
          \Big \{ \bar{\bbox{\Delta}}_\mu
         \bbox{\cdot} {\rm tr}[i\partial^\mu  (U - U^\dagger )
            \bbox{\tau} ]  N
             + \bar{N} {\rm tr}[i\partial^\mu  (U - U^\dagger )
            \bbox{\tau} ] \bbox{\cdot \Delta}_\mu \Big\}
            +\cdots
                   \ .     \label{eq:L4}
\end{eqnarray}
Here we have used the definitions
\begin{eqnarray}
 \stackrel{\leftrightarrow}{\cal D}_\mu
     &=& {\cal D}_\mu -
        (\stackrel{\leftarrow}{\partial}_\mu - iv_{\mu}) \quad;\quad
v_{\mu\nu} = \partial_{\mu} v_{\nu} -\partial_{\nu}
      v_{\mu} + i [v_{\mu}, v_{\nu}] 
       = - i [a_\mu, a_\nu]   \ , \\[4pt]
    D_\mu a_{\nu} &=& \partial_\mu a_{\nu}
                            + i [v_\mu, a_{\nu}]  \quad;\quad
    D_\sigma v_{\mu\nu} = \partial_\sigma v_{\mu\nu}
                            + i [v_\sigma, v_{\mu\nu}] \ ,
\end{eqnarray}

In Eqs.~(\ref{eq:L3}) and (\ref{eq:L4}) the ellipsis represents terms 
that do not contribute to the $\pi N$ scattering amplitude. 

\subsection{Scattering Amplitudes}

\begin{figure}
 \setlength{\epsfxsize}{2.5in}
  \centerline{\epsffile{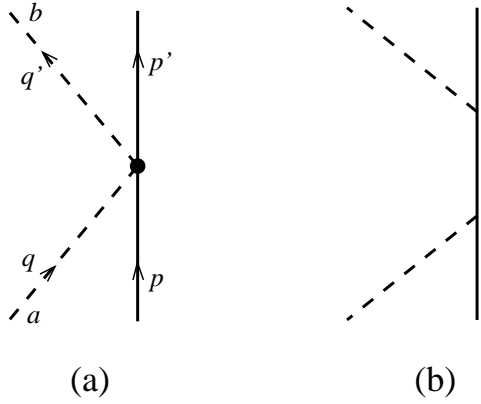}}
\vspace*{.2in}
\caption{Tree-level diagrams for $\pi N$ scattering: (a) contact 
interactions; (b) nucleon exchange with the cross diagram suppressed}
 \label{fig:tree}
\vspace*{.4in}
\end{figure}

Following the 
standard notation of H\"{o}hler \cite{HOHLER83} for $\pi N$ scattering
we  write  the $T$ matrix as
\begin{equation}
  T_{ba} \equiv \langle \pi_b | T | \pi_a\rangle
         =T^+ \delta_{ab} + \case{1}{2}[\tau_b, \tau_a] T^- \ ,
\end{equation}
where the isospin symmetric and antisymmetric amplitudes are 
\begin{equation}
    T^{\pm} = \bar{u}(p')\left[A^{\pm} + \case{1}{2}(\rlap / q +
\rlap/ q') B^\pm\right]u(p) \ ,
\end{equation}
and $u$ denotes a nucleon spinor.
Here, as shown in Fig. \ref{fig:tree}, $q$ and $q'$ are 
the c.m. momenta of the incoming and 
outgoing pions with isospin labels $a$ and $b$ respectively. The
c.m. momenta of the incoming and outgoing nucleons are labelled
$p$ and $p'$ respectively. The amplitudes
$A^\pm$ and $B^\pm$ are functions of the Mandelstam 
invariants $s=(p+q)^2$, $t=(q-q')^2$, and $u=(p-q')^2$.

In Fig. \ref{fig:tree} we show the tree level Feynman diagrams
arising from the contact terms and from one nucleon exchange.
The vertex in Fig. \ref{fig:tree}(a) arises from any of the 
interactions in ${\cal L}_3$ and ${\cal L}_4$ (except for the 
$\lambda_1$ and $\lambda_6$ terms), as well as the Weinberg term
$-\bar{N}\gamma^{\mu}v_{\mu}N$ of ${\cal L}_2$. 
The amplitudes for these tree diagrams were given in I.

\subsubsection{$\Delta$ Exchange}

\begin{figure}
 \setlength{\epsfxsize}{3in}
  \centerline{\epsffile{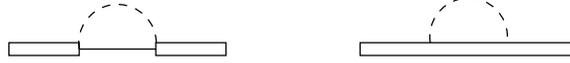}}
\vspace*{.2in}
\caption{One-loop $\Delta$ self-energy diagrams. The 
open box represents the free $\Delta$ propagator.}
 \label{fig:seven}
\vspace*{.4in}
\end{figure}

When the $\Delta$ appears as an intermediate state for $\pi N$ 
scattering, the tree-level $T$-matrix diverges at $s = M_\Delta^2$. 
In I we argued that the power counting should work only for irreducible 
diagrams. Therefore we summed diagrams containing the one-particle 
irreducible self-energy insertions shown in Fig.~\ref{fig:seven} 
to all orders so as to replace the free propagator by the
dressed propagator which is finite. To \ord{3} this gave for the real 
part of the $\Delta$ propagator in $d$ dimensions 
\begin{eqnarray}
\Re G_{\mu\nu}(k) &=& -\Delta_R(k^2)
(\rlap/{\mkern-1mu} k + M_\Delta)(P^{3/2})_{\mu\nu}
             -\frac{1}{\sqrt{d-1}M_{\Delta}}
             (P^{1/2}_{12}+P^{1/2}_{21})_{\mu\nu}\nonumber\\
&&\qquad\qquad+\frac{(d-2)}{(d-1)M_{\Delta}^2}(\rlap/{\mkern-1mu} k +M_{\Delta})
             (P^{1/2}_{22})_{\mu\nu} \ ,
                      \label{eq:fullGmunu}
\end{eqnarray}
where the spin projection operators \cite{NIEU81,BEN89}, denoted by
$\left(P_{ij}^I\right)_{\mu\nu}$, are generalized by replacing factors of 2 
and 3 by $(d-2)$ and $(d-1)$, respectively. In Eq. (\ref{eq:fullGmunu}) we 
have defined
\begin{equation}
 \Delta_R(k^2) = {k^2- M_\Delta^2 - \Pi_\Delta(k^2) \over
               [k^2- M_\Delta^2 - \Pi_\Delta(k^2)]^2 +
                    M_\Delta^2\Gamma_\Delta^2(k^2) }\;.
\end{equation}
(Obviously the bare propagator corresponds to setting $\Pi_\Delta$ and 
$\Gamma_\Delta$ to zero.)

The diagrams which contribute to the self energy are shown in 
Fig. \ref{fig:seven}, where the open box denotes a free $\Delta$ 
propagator. Only the $g_{\mu\nu}$ part is needed in leading order
and the renormalized self energy is
\begin{equation}
\Sigma_\Delta^{\rm ren}(k) =\Sigma_\Delta(k) 
-\Re\Sigma_\Delta(k)\Bigm|_{\rlap\slat{\mkern-1mu}k=M_\Delta}
-{\partial \over \partial\rlap/{\mkern-1mu}k}
\Re\Sigma_\Delta(k)\Bigm|_{\rlap\slat{\mkern-1mu} k=M_\Delta}
            (\rlap/{\mkern-1mu} k-M_\Delta)\;.
\end{equation}
The integrals which arise are 
evaluated in the IR scheme, which is briefly discussed in the Appendix. 
For the $\gamma$-matrix algebra we evaluate terms whose leading 
contribution is at \ord{3} and we discard terms for which the leading
contribution is of higher order. We include polynomial 
terms obtained from the product of a $1/(d-4)$ singularity with $(d-4)$ 
factors which were were dropped in I. Following this 
procedure the real part of the $\Delta$ polarization in the $s$- or 
$u$-channel is given by
\begin{eqnarray}
\Pi_{\Delta}(s)&=&-\frac{h_A^2}{(4\pi f_{\pi})^2}\biggl\{(s+2MM_{\Delta}
+M_{\Delta}^2)\Bigl[\bar{\cal I}_{11}^{(2)}(s,M)-
\bar{\cal I}_{11}^{(2)}(M_\Delta^2,M)\Bigr]\nonumber\\
&&\qquad-2(s-M_{\Delta}^2)M_{\Delta}(M_{\Delta}+M)
\frac{\partial}{\partial M_{\Delta}^2}
\bar{\cal I}_{11}^{(2)}(M_\Delta^2,M)\biggr\}\nonumber\\
&&-\frac{5\tilde{h}_A^2}{36(4\pi f_{\pi})^2}\biggl\{2(s+9M_{\Delta}^2)
\Bigl[\bar{\cal I}_{11}^{(2)}(s,M_{\Delta})-
\bar{\cal I}_{11}^{(2)}(M_{\Delta}^2,M_{\Delta})\Bigr]\nonumber\\
&&\qquad-20(s-M_{\Delta}^2)M_{\Delta}^2\frac{\partial}{\partial k^2}
\bar{\cal I}_{11}^{(2)}(k^2,M_\Delta)\biggl|_{k^2=M_{\Delta}^2}
\nonumber\\
&&+\frac{13(s-M_{\Delta}^2)}{36s^2}\biggl((s+3M_{\Delta}^2)
\Bigl[6sm^2-(s-M_{\Delta}^2)^2\Bigr]-24s^2m^2\biggr)\biggr\}
\,,\label{jkp}
\end{eqnarray}
where the integral $\bar{\cal I}_{11}^{(2)}$ is specified in the Appendix.
By suitably approximating the above expression the HB
form given in I is obtained (apart from polynomial terms).
The width is zero in the $u$-channel and in the $s$-channel it is
\begin{equation}
\Gamma_\Delta(s)=\frac{\pi h_A^2}{12M_\Delta s^2(4\pi f_{\pi})^2}
\left[(s-M^2+m^2)^2-4sm^2\right]^{\frac{3}{2}}(s+M^2+2MM_\Delta-m^2)\;,
\label{delwidth}
\end{equation}
which gives the exact result \cite{HOHLER83} on shell at $s=M_\Delta^2$.

\begin{figure}
 \setlength{\epsfxsize}{1in}
  \centerline{\epsffile{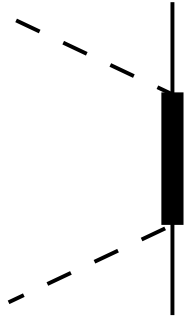}}
\vspace*{.2in}
\caption{The $\Delta$ exchange diagram with the dressed propagator
represented by a solid box. The cross diagram is not shown.}
 \label{fig:delexch}
\vspace*{.4in}
\end{figure}

The $\Delta$ exchange tree-level diagram for $\pi N$ scattering is 
pictured in Fig. \ref{fig:delexch}, where the solid box denotes the 
dressed propagator discussed above (a similar notation is used in 
Fig. \ref{fig:vtx} below). The corresponding contribution to the 
$T$-matrix can be obtained from the expression given in I. 

Including also the loop diagrams, discussed below, 
the real part of the total $T$-matrix yields the real part of the elastic 
scattering amplitude, $f_\alpha$, by means of 
the standard partial wave expansion\cite{GASIO66}. 
Here the isospin-spin partial wave channels are labelled by
$\alpha\equiv(l,2I,2J)$ with $l$ the orbital angular momentum,
$I$ the total isospin, and $J=l\pm{1\over 2}$ the total angular momentum.
The phase shifts $\delta_\alpha$ are then given by 
\begin{equation}
     \Re f_\alpha = {1\over |\bbox{q}|}
            \Re e^{i\delta_\alpha}\sin \delta_\alpha  \;,   \label{eq:f}
\end{equation}
where $|\bbox{q}|$ is the magnitude of the c.m. three-momentum.
We shall refer to this as the ``$S$-matrix" approach.

An alternative, which has been espoused by Fettes and Mei\ss ner
\cite{fm3nd,fm3n,fm4}, is the $K$-matrix approach. 
This is introduced by setting
\begin{equation}
f_\alpha = {K_\alpha \over 1 - i |\bbox{q}| K_\alpha}
\quad{\rm where}\quad     
K_\alpha = {1\over |\bbox{q}|} \tan\delta_\alpha \ . \label{eq:Kf}
\end{equation}
The calculated amplitude $\Re f_\alpha$ is then assumed to actually 
be $K_\alpha$ so that an infinite $f_\alpha$ on resonance corresponds to 
$\delta=\thalf\pi$ (for further discussion see \cite{EW88,ELLIS96}).
This allows the free $\Delta$ propagator to be used everywhere.
Thus it is not necessary to sum the self energy insertions of Fig.
\ref{fig:seven} and we can expand $\Delta_R$ to first order in 
$\Pi_\Delta$. The zeroth order term gives the tree diagram of 
Fig. \ref{fig:delexch}, but with the free propagator, and the first order term
is treated in the same way as the other third order loop diagrams.
We will compare this method with the $S$-matrix approach.

\subsubsection{One-Loop Diagrams}

\begin{figure}
 \setlength{\epsfxsize}{5in}
  \centerline{\epsffile{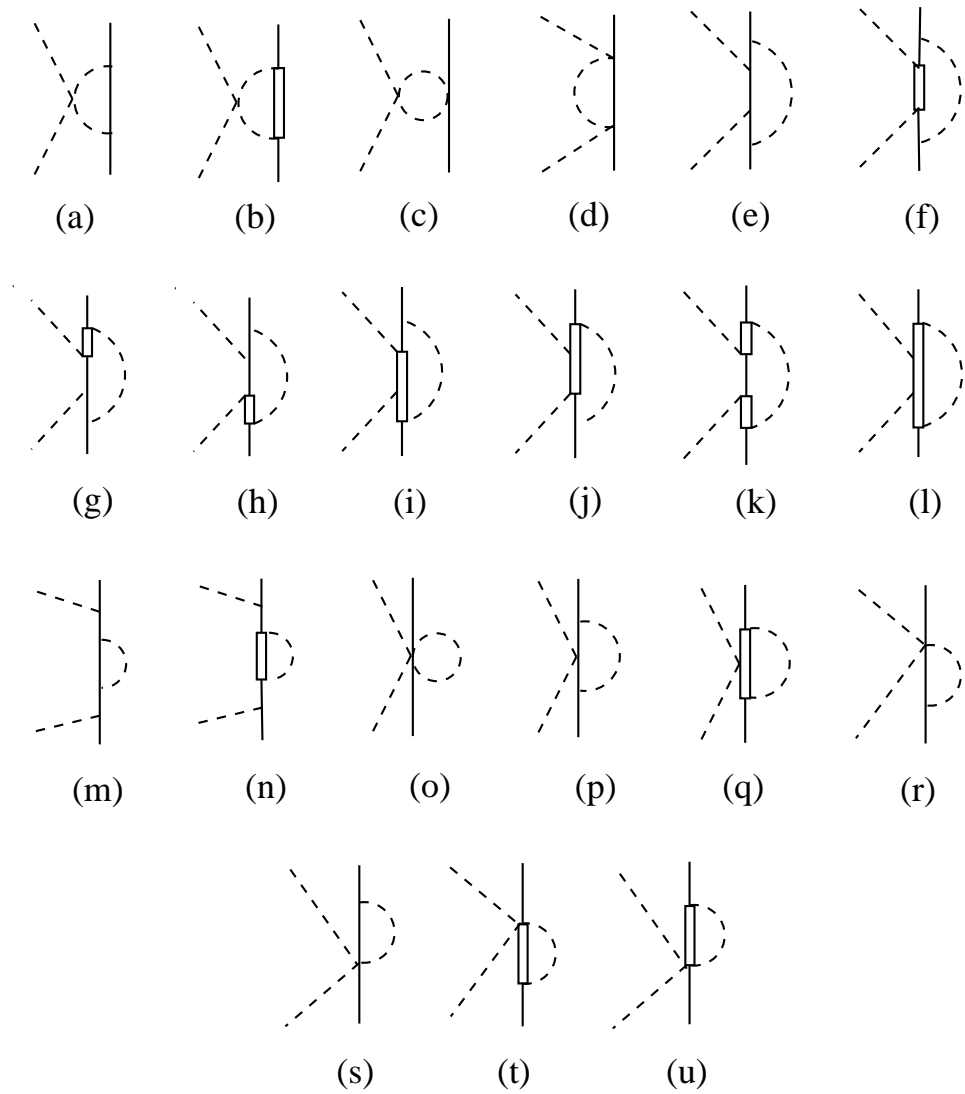}}
\vspace*{.2in}
\caption{A set of one-loop diagrams which contribute at \ord{3}. 
Crossed diagrams for (d) to (n) are not shown. }
 \label{fig:nonvtx}
\vspace*{.4in}
\end{figure}

A set of one-loop diagrams that contribute to the $\pi N$ $T$-matrix at
\ord{3} is shown in Fig.~\ref{fig:nonvtx}. Here we can use
the free $\Delta$ propagator, denoted by an open box, since no 
singularities are generated in the $T$-matrix. We 
illustrate our procedure by discussing the evaluation of 
Fig.~\ref{fig:nonvtx}(g), which has the same value as 
Fig.~\ref{fig:nonvtx}(h). Using the standard Feynman rules, we find 
\begin{eqnarray}
T_{4g}&=&-\bar{u}(p')\frac{g_A^2h_A^2}{4f_\pi^4}i\mu^{4-d}\int_I
\frac{d^d\ell}{(2\pi)^d}\ell^\nu T^cG_{\nu\mu}(p'+\ell)T^{\dagger b} 
q^{\prime\mu}
\nonumber\\
&&\qquad\times\frac{1}{\rlap/{\mkern-1mu}p+\rlap/q 
+\rlap/{\mkern0.75mu}\ell-M+i\epsilon}
\rlap/q\gamma_5\tau^a\frac{1}{\rlap/{\mkern-1mu}p
+\rlap/{\mkern0.75mu}\ell-M+i\epsilon}\rlap/{\mkern0.75mu}\ell
\gamma_5\tau^c\frac{1}{\ell^2-m^2+i\epsilon}u(p)\;,
\end{eqnarray}
where $\mu$ is the renormalization scale, $T^i$ denotes a component 
of the isospin $\thrhalf$ to $\thalf$ transition matrix and the subscript 
$I$ denotes that the integral is to be evaluated using infrared 
regularization. Rather than using the projection operators in 
(\ref{eq:fullGmunu}) it is more convenient to write the free 
$\Delta$ propagator in the form
\begin{equation}
G_{\nu\mu}(k)=\frac{1}{\rlap/{\mkern-1mu}k-M_\Delta+i\epsilon}
\left[ -g_{\nu\mu}+\frac{1}{d-1}\gamma_\nu\gamma_\mu
+\frac{1}{(d-1)M_\Delta}(\gamma_\nu k_\mu- 
k_\nu\gamma_\mu)+\frac{(d-2)}{(d-1)M_\Delta^2}k_\nu k_\mu \right].
\end{equation}
To \ord{3} the $\rlap/{\mkern0.75mu}\ell$ terms in the baryon denominators 
can be dropped. Then the integral involves $\ell^\nu\ell^\rho$ in the 
numerator and to this order only the $g_{\nu\rho}$ contribution is needed.
Carrying out the $\gamma$-matrix and isospin algebra, we obtain
\begin{eqnarray}
T_{4g}&=&-\bar{u}(p')\thalf[\tau^b,\tau^a]\frac{4g_A^2h_A^2}
{9f_\pi^2(4\pi f_\pi)^2}
M^2(M+M_\Delta)(s-M^2-m^2+\thalf t-2M\rlap/q)\nonumber\\
&&\times\left[\bar{\cal I}_{13}^{(3)}(s,t,M_\Delta,M,M)
-\frac{1}{24M^4}(s-M_\Delta^2-m^2)
\left(\tfourth+32\pi^2L+\ln\frac{m^2}{\mu^2}\right)\right]u(p)\;.
\end{eqnarray}
The integral $\bar{\cal I}_{13}^{(3)}$ is defined in the Appendix, as
is the divergent quantity $L$. In the usual way the terms involving
$\left(2L+\frac{1}{(4\pi)^2}\ln\frac{m^2}{\mu^2}\right)$ can be 
absorbed in the low energy constants so that the result is independant 
of the renormalization scale. This expression can be reduced to HB 
form. Here, as in I, we define the average baryon 
mass, $\bar{M}=\thalf(M+M_\Delta)$ and $\eta=(s-M^2)/(2\bar{M})$. 
Then evaluating the expressions to leading order and using the reduction 
of $\bar{\cal I}_{13}^{(3)}$ given in the Appendix, we obtain
\begin{equation}
T_{4g}\rightarrow\bar{u}(p')\thalf[\tau^b,\tau^a]\frac{4g_A^2h_A^2}
{27f_\pi^2(4\pi f_\pi)^2}\left[M\rlap/q-\bar{M}\eta
+\tquar(2m^2-t)\right]
\left[F_2(\eta,-\delta)-\ttwoth(\eta-\delta)\right]u(p),
\end{equation}
recalling that $\delta$ denotes the delta-nucleon mass difference.
This agrees with the result given in I, apart from the polynomial term 
which we include here, but which was subsumed in the low energy constants in 
I. To these results in the $s$-channel should be added the contribution in 
the $u$-channel with the replacement $s\rightarrow u$ and the interchanges
$a\leftrightarrow b$ and $q\leftrightarrow -q'$.

\begin{figure}
 \setlength{\epsfxsize}{4.5in}
  \centerline{\epsffile{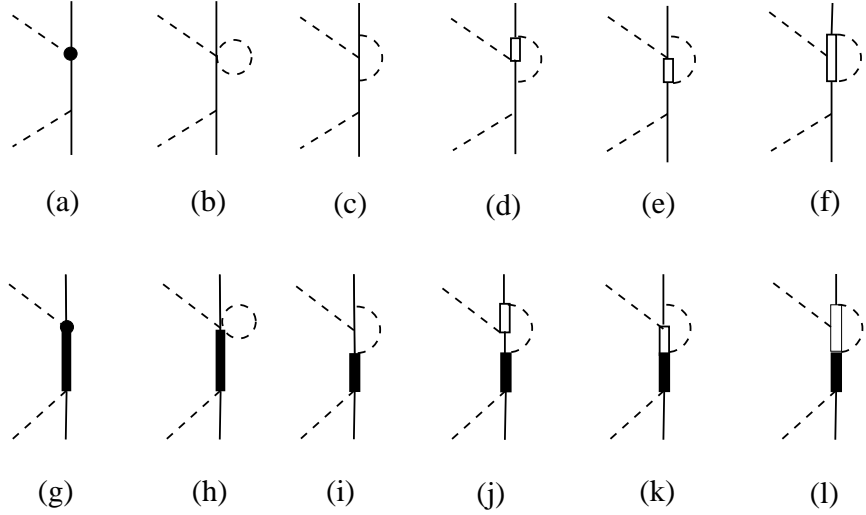}}
\vspace*{.4in}
 \caption{Diagrams with one-loop vertices which contribute at 
\ord{3}. Crossed diagrams are not shown. The solid circle in (a) and (g)
refers to $\lambda_1$ and $\lambda_6$ vertices, respectively.
Each diagram implicitly includes its counterpart
where the lower vertex is dressed.}
\vspace*{.4in}
 \label{fig:vtx}
\end{figure}

Proceeding in similar fashion the real parts of the $T$-matrix for the 
diagrams of Fig.~\ref{fig:nonvtx} may be calculated.
Note that the diagrams in  
Fig.~\ref{fig:nonvtx}(m) and (n) involve the nucleon one-loop 
self-energy which is real for the energies of interest here. 
As with the $\Delta$, we make on-shell mass and wavefunction 
counterterm subtractions to obtain the renormalized self-energy,
namely
\begin{equation}
 \Sigma_N^{\rm ren}(k) =\Sigma_N(k) 
-\Sigma_N(k)\Bigm|_{\rlap\slat{\mkern-1mu} k=M}
-{\partial \over \partial  \rlap/{\mkern-1mu} k }
\Sigma_N(k)\Bigm|_{\rlap\slat{\mkern-1mu} k=M}
(\rlap/{\mkern-1mu} k-M)\;. \label{eq:subnself}
\end{equation}
Since these diagrams do not give singular contributions to the $T$-matrix
we do not sum these self-energy insertions. 

We also need to evaluate in similar fashion the diagrams of 
Fig.~\ref{fig:vtx}, where the solid boxes denote the 
dressed propagator discussed in Subsec. II.B.1. The diagrams of 
Fig.~\ref{fig:vtx}(a)--(f) modify the $\pi NN$ tree vertex in 
${\cal L}_2$, while diagrams (g)--(l) similarly modify the 
$\pi N\Delta$ tree vertex. 

In  a few cases the IR results can be checked against Ref. 
\cite{blpi}. The reduction of the IR expressions to HB form also provides 
a check since the results of I should be reproduced (these, in turn, 
were checked against Moj\v{z}i\v{s} \cite{MO97} for the cases without a 
$\Delta$). In the course of checking we found a phase error in one of the 
HB results in I. This means that the expression given in I for diagram 
\ref{fig:nonvtx}(i), which is equal to that for diagram  
\ref{fig:nonvtx}(j), should be multiplied by a factor of 5. This has
little impact on the fits presented in I, but the parameters are modified
somewhat. Since the IR 
results will not prove to be satisfactory, we shall not list the rather 
lengthy expressions for all the diagrams; they are available on request.

\subsection{$\sigma$ Term and Effective Couplings}

We may obtain the nucleon $\sigma$ term from the Feynman-Hellman
theorem,
\begin{equation}
  \sigma(0) = m^2 {\partial M \over \partial m^2 } \ .
\end{equation}
The nucleon mass receives contributions from the $\kappa_2$ term in ${\cal 
L}_3$ and diagrams (m) and (n) of Fig.~\ref{fig:nonvtx}. In the IR scheme
these yield
\begin{eqnarray}
\sigma(0)&=&-\frac{4\kappa_2m^2}{M}-\frac{9g_A^2Mm^2}{4(4\pi f_\pi)^2}
\bar{\cal I}_{11}(M^2,M)\nonumber\\
&&\qquad-\frac{2h_A^2(M+M_\Delta)m^2}{(4\pi f_\pi)^2}\left[
\bar{\cal I}_{11}(M^2,M_\Delta)+\frac{(M^2-M_\Delta^2)(M^2+2M_\Delta^2)}
{6M^2M_\Delta^2}\right]\;.
\end{eqnarray}
In the HB approximation this reduces to
\begin{equation}
\sigma(0)\rightarrow-\frac{4\kappa_2m^2}{M}-\frac{9\pi g_A^2m^3}
{4(4\pi f_\pi)^2}
-\frac{4h_A^2m^2}{(4\pi f_\pi)^2}[J(\delta)-\delta]\;,
\end{equation}
where the integral $J$ is defined in I, see also the Appendix.
This agrees with the result of Fettes and Mei\ss ner \cite{fm3nd} and, 
modulo a polynomial contribution in $m^2\delta$, with I.

The $\pi NN$ vertex up to one-loop order consists of the tree vertex generated
from the axial $a_\mu$ term in ${\cal L}_2$ and 
the one-loop diagrams shown in the upper part of 
Fig.~\ref{fig:vtx} -- diagrams (a) to (f).
As in I, we can calculate the 
one-loop vertex function $\Gamma^a(k,k',q)$, where $k$($k'$) is the 
incoming (outgoing) momentum of the nucleon and $q=k'-k$ is the momentum
transfer. The $\pi NN $ coupling for on-shell nucleons is then obtained 
from
\begin{equation}
    \bar{u}(k') \Gamma^a(k,k',q) u(k)
          = g_{\pi NN}(q^2)  \bar{u}(k')\gamma_5 \tau^a u(k) \ .
\end{equation}
At zero momentum transfer we obtain, to \ord{2},
\begin{eqnarray}
\zeta_{\pi NN}&\equiv&\frac{g_{\pi NN}(0)f_\pi}{M g_A}=
1-\frac{2m^2\lambda_1}{g_AM^2}+\frac{g_A^2m^2}{2(4\pi f_\pi)^2}
+\frac{64h_A^2m^2M}{27(M_\Delta-M)(4\pi f_\pi)^2}\nonumber\\
&&\qquad\times\Biggl\{\bar{\cal I}_{11}(M^2,M)
+\frac{[(M^2-M_\Delta^2)^2-2m^2(M^2+M_\Delta^2)]}{4m^2M^2}
\bar{\cal I}_{11}(M^2,M_\Delta)\nonumber\\
&&\qquad\qquad\qquad+\frac{(M_\Delta^2-M^2)}{24m^2M^4}
[(M^2-M_\Delta^2)^2-6m^2M^2]\Biggr\}\nonumber\\
&&-\frac{10h_A^2\tilde{h}_A(M+M_\Delta)}
{81g_AM_\Delta^2(4\pi f_\pi)^2}
\biggl\{\left[7MM_\Delta^2+8M_\Delta^3-M^2(2M+3M_\Delta)\right]
\bar{\cal I}_{11}^{(1)}(M^2,M_\Delta) \nonumber\\
&&+\frac{1}{12M^4}[35MM_\Delta^2
+37M_\Delta^3-2M^2(2M+3M_\Delta)][2m^2M^2-(M^2-M_\Delta^2)^2]\biggr\}
\ .  \label{eq:gpiNN}
\end{eqnarray}
With the parameters obtained 
from fits to the $\pi N$ phase shifts this 
allows a test of the Goldberger-Treiman relation.

Similarly, we can calculate the $\pi N \Delta$ vertex
from the tree-level $h_A$ term
and the one-loop diagrams shown in the lower part of 
Fig.~\ref{fig:vtx} -- diagrams (g) to (l).
Here, in the vertex function $\Gamma^{\mu a}(k,k',q)$ the label
$k$ now refers to the incoming $\Delta$ momentum.
The $\pi N\Delta$ coupling is obtained from
\begin{equation}
    \bar{u}(k') \Gamma^{\mu a}(k,k',q) u_\mu (k)
= g_{\pi N\Delta}(q^2)  \bar{u}(k')q^\mu T^a u_\mu (k) \ ,
\end{equation}
where $u_\mu(k)$ is the $\Delta$ spinor. At zero momentum 
transfer we obtain, to \ord{2},
\begin{eqnarray}
\frac{g_{\pi N\Delta}(0)f_\pi}{M h_A} &=& 1
-\frac{2m^2\lambda_6}{h_AM^2}-\frac{4g_A^2M^2}{(4\pi f_\pi)^2}
\bar{\cal I}_{12}^{(3)}(M_\Delta^2,M,M)\Bigm|_{m=0}
-\frac{65g_A\tilde{h}_A(M+M_\Delta)m^2}{108M(4\pi f_\pi)^2}
\nonumber\\
&&-\frac{2h_A^2M(M+M_\Delta)}{9(4\pi f_\pi)^2}\Biggl\{
\bar{\cal I}_{12}^{(3)}(M_\Delta^2,M,M_\Delta)\Bigm|_{m=0}
+\frac{\left[(M_\Delta^2-M^2)^2-6m^2M^2\right]}{18M^4}\Biggr\}
\nonumber\\
&&-\frac{50\tilde{h}_A^2M_\Delta(M+M_\Delta)}{81(4\pi f_\pi)^2}
\Biggl\{\bar{\cal I}_{12}^{(3)}
(M_\Delta^2,M_\Delta,M_\Delta)\Bigm|_{m=0}\nonumber\\
&&\qquad\qquad\qquad\qquad
+\frac{(15M+M_\Delta)[(M_\Delta^2-M^2)^2-6m^2M^2]}{360M^4M_\Delta}
\Biggr\}\;. \label{eq:gpiNdelta}
\end{eqnarray}
Note that for the first two integrals $\bar{\cal I}_{12}^{(3)}$ both 
real and imaginary parts need to be considered.
The value of $g_{\pi N\Delta}$ is complex because the 
intermediate pion and nucleon states for Fig.~\ref{fig:vtx}(i) 
and (j) can go on shell. We therefore define
$\zeta_{\pi N\Delta}\equiv|g_{\pi N\Delta}(0)f_\pi/(M h_A)|$.
The HB reduction of Eqs. (\ref{eq:gpiNN}) and  (\ref{eq:gpiNdelta})
agrees with the expressions in I up to polynomial terms.

\section{Results}

As fixed input parameters, we use the standard baryon and pion masses:
$M=939\,$MeV, $M_\Delta=1232\,$MeV, and $m=139\,$MeV.
We also take\cite{PDT} $f_\pi = 92.4\,$MeV from charged pion decay, 
$g_{\rm A} = 1.26$ from
neutron $\beta$ decay, and $h_{\rm A}=1.46$ from 
Eq.~(\ref{delwidth}) for $\Gamma_\Delta(M_\Delta^2)$ using the central 
value of the $\Delta$ width, $\Gamma_\Delta = 120\pm5\,$MeV.
We have ten low energy constants: $\beta_\pi$, $\kappa_\pi$,
$\kappa_1$, $\kappa_2$, and $\lambda_1$ to $\lambda_6$, plus
the $\pi\Delta\Delta$ coupling, $\tilde{h}_A$.
These are obtained by optimizing 
the fit of our calculated $\pi N$ $S$- and $P$-wave phase
shifts to the 2001 data of the VPI/GW group \cite{ARNDT}.
Since errors are not given we compared results obtained
by assigning all the data points the same relative weight
to those obtained using weightings suggested in the literature
\cite{fm3nd,ulol}. The differences were small so we used the same relative 
weightings, minimizing
\begin{equation}
\chi^2=\frac{1}{N}\sum_{i=1}^N\left(
\frac{\xi_i^{\rm theory}-\xi_i^{\rm experiment}}{0.08\xi_i^{\rm experiment}}
\right)^{\!2}\;,
\end{equation}
where $\xi_i=\sin2\delta_i$ for the $S$-matrix calculation and 
$\xi_i=2\tan\delta_i$ for the $K$-matrix calculation.

\subsection{HB Results}

\begin{figure}[p]
\setlength{\epsfysize}{7.9in}
\centerline{\epsffile{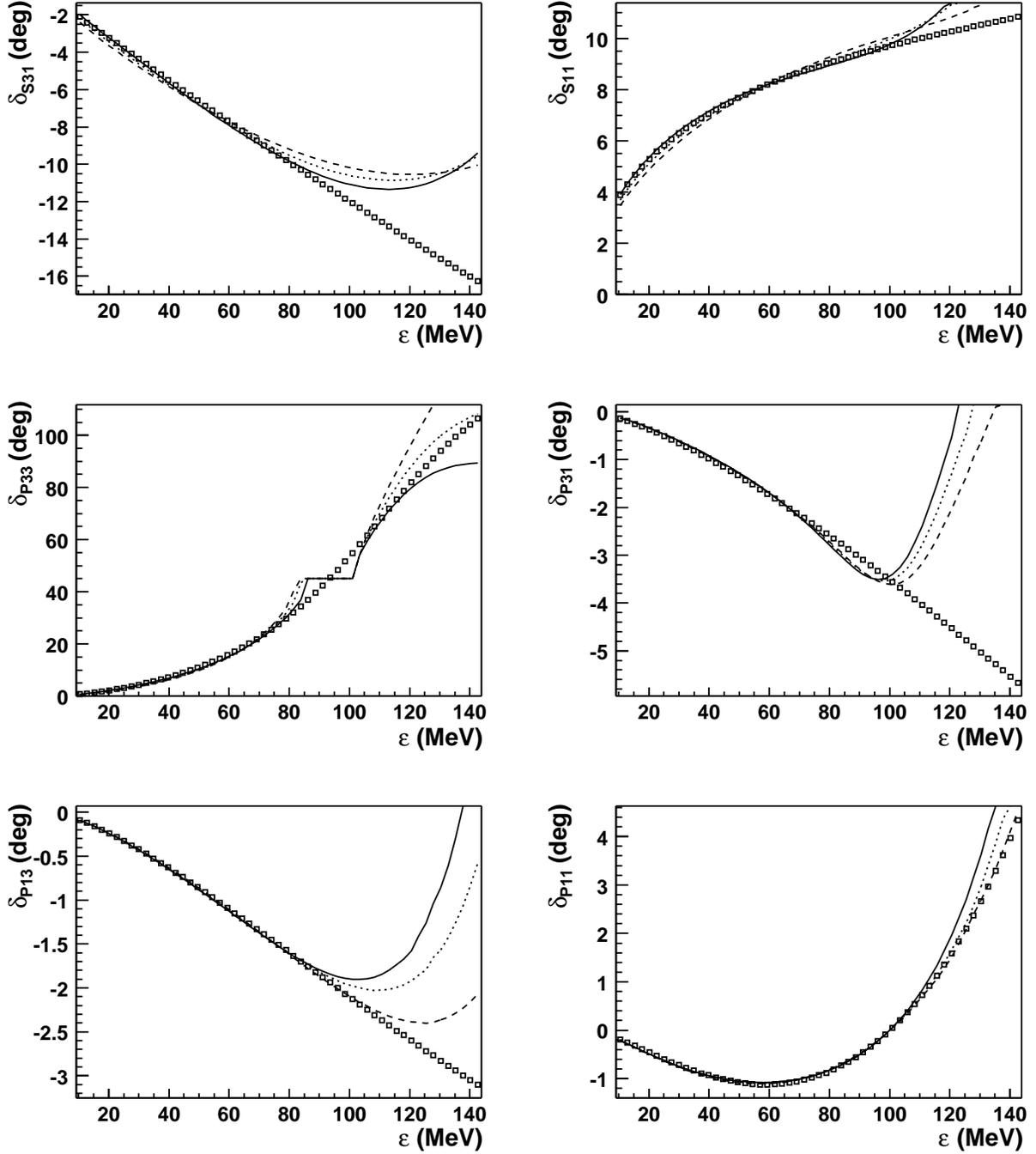}}
\caption{The $S$- and $P$-wave phase shifts using the $S$-matrix method 
as a function of the pion c.m. kinetic energy, $\epsilon$. The solid curve 
corresponds to an unconstrained nucleon $\sigma$ term, while for the dotted 
and dashed curves it is constrained to be 75 MeV and 45 MeV, respectively. 
The data are from Ref. \protect\cite{ARNDT}}
\label{hbs}
\end{figure}
\begin{figure}[p]
\setlength{\epsfysize}{7.9in}
\centerline{\epsffile{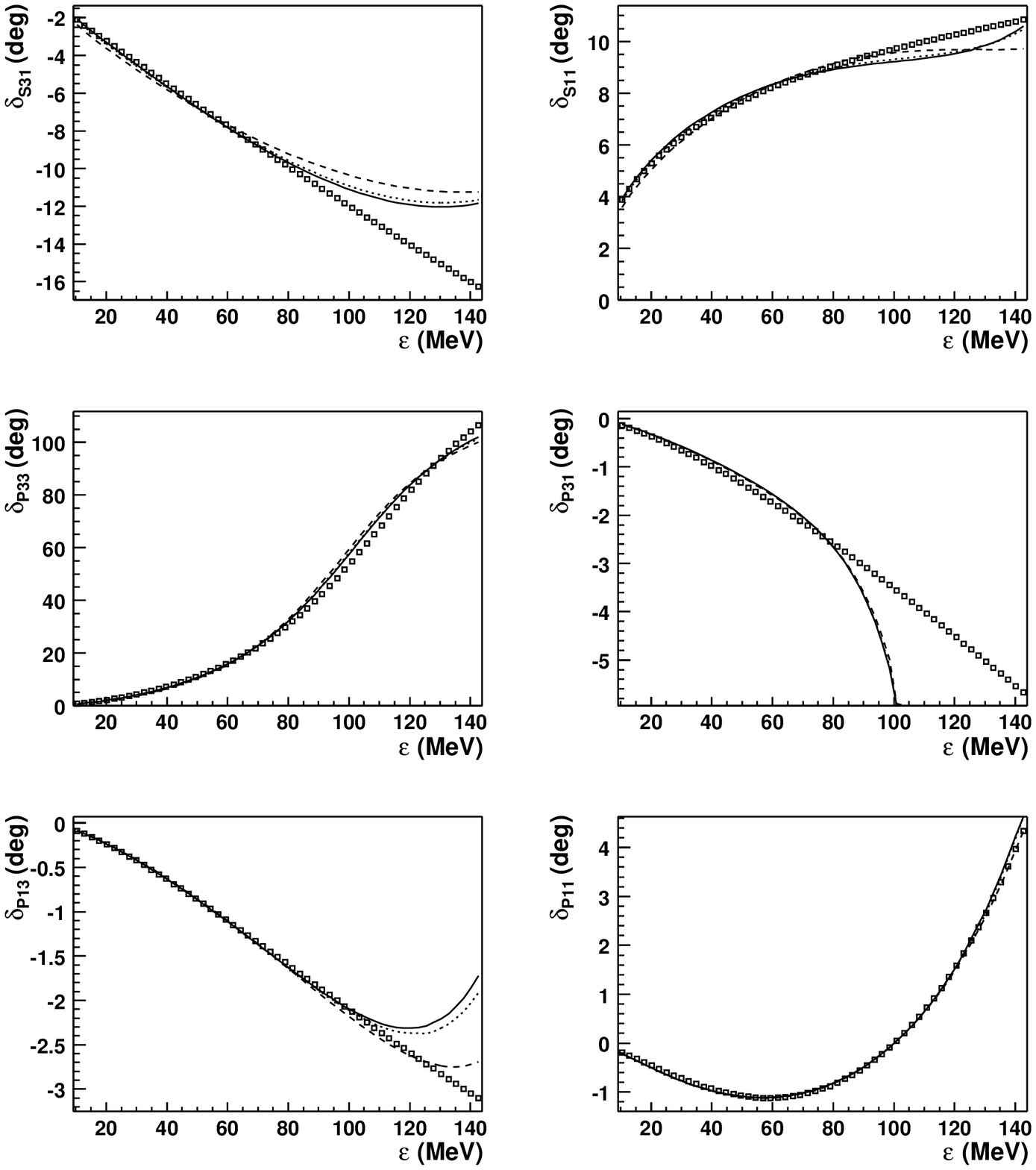}}
\caption{As for Fig. \protect\ref{hbs}, but using the $K$-matrix method.}
\label{hbk}
\end{figure}

Our fits to the $S$- and $P$-wave phase shifts in the HB calculation are 
shown  in Figs. \ref{hbs} and \ref{hbk} using the $S$-matrix and $K$-matrix 
methods, respectively. Here the solid curves correspond to an unconstrained 
fit. As in I and Ref. \cite{fm3nd} it was possible to fit the phase shift 
data out to a pion c.m. kinetic energy, $\epsilon$, of 100 MeV, slightly 
below the $\Delta$ resonance at $\epsilon=127$ MeV. More precisely the data 
were fitted from $\sqrt{s}=1090$ to 1200 MeV in 3 MeV steps.
The short horizontal section in 
Fig. \ref{hbs} indicates a small region where unitarity, which is only
enforced perturbatively, is slightly violated 
and a phase shift cannot be determined; this, of course, does not occur in 
the $K$-matrix approach, by definition. The corresponding parameters are 
collected in Table \ref{tab1} where, in common with previous work, it is 
observed that the deduced values
of the $\sigma$ term are much larger than the oft-quoted figure 
of $45\pm 8$ due to Gasser et al. \cite{gas}. Therefore we performed 
two additional fits where $\kappa_2$ was constrained to produce 
values of the $\sigma$ term of 75 MeV and 45 MeV. These are indicated, 
respectively, by the dotted and dashed curves in Figs. \ref{hbs} and 
\ref{hbk}. For the $K$-matrix case the dotted curve is barely 
distinuishable from the solid curve since the change in the $\sigma$
term is small here.

As expected the $S$-matrix fits in Fig. \ref{hbs} are quite similar to
those given in I. The fitted parameters, however, differ since here we 
include the polynomial terms and we subsume the renormalization scale 
dependance into the low energy constants. In many of the partial waves 
the $K$-matrix fits (Fig. \ref{hbk}) are a little better, particularly 
for $\delta_{P33}$, however $\delta_{P31}$ begins to deviate significantly
from the data at $\epsilon\sim80$ MeV. Something of the same trend is 
visible in the work of Fettes and Mei\ss ner \cite{fm3nd}. The 
\ord{2} low energy constants given there are related to our constants by
$c_1=\kappa_2/M$, $c_2=\kappa_1/(2M)$, $c_3=2\beta_\pi/M$ and 
$c_4=\kappa_\pi/M$. Their fit $1^\dagger$ to Matsinos' data \cite{mat} produces 
values which are fairly similar to ours when the different value of the 
$\pi N\Delta$ coupling, $h_A$, and the constraint $\sigma(0)=59$ MeV are
taken into account. (We should note, however, that substantial differences 
are seen for their fit to the Karlsruhe data \cite{karl}.) The \ord{2} 
constants in Table \ref{tab1} are reasonably similar in the $S$- and 
$K$-matrix calculations, however sizeable differences appear at \ord{3}.
Further they are not of natural size (unity). These features may be due to 
the strong cancellations which occur among the various terms and the fact 
that the parameters are not uniquely determined since it is possible to 
obtain similar values of $\chi^2$ when $\sigma(0)$ is constrained. A 
significantly larger 
$\chi^2$ is only obtained for the $S$-matrix case with $\sigma(0)=45$ MeV.

\begin{table}
\caption{Low energy constants and effective couplings obtained from the 
$S$- and $K$-matrix fits with
unconstrained and constrained values of the nucleon $\sigma$ term.}
\begin{center}
\begin{tabular}{|r|rrr|rrr|}
&\multicolumn{3}{c|}{$S$-matrix}&\multicolumn{3}{c|}{$K$-matrix}\\ 
	\hline
$\beta_{\pi}$ &   4.78 &   4.46 &   4.13 &   4.06 &   4.06 &   4.05 \\
$\kappa_{\pi}$& $-$10.1 &  $-$9.02 &  $-$7.76 &  $-$7.96 &  $-$7.79 
&  $-$7.21 \\
$\kappa_1$    & $-$10.6 &  $-$9.80 &  $-$8.82 &  $-$9.36 &  $-$9.29 
&  $-$9.04 \\
$\kappa_2$    &  $-$0.453 &  $-$0.124$^\ast$  &  0.241$^\ast$  
		 &  $-$0.234 &  $-$0.124$^\ast$ &   0.240$^\ast$ \\
$\lambda_1$   &  23.7 &  16.6 &   8.20 &   9.11 &   7.77 &   3.13 \\
$\lambda_2$   & $-$20.1 & $-$18.0 & $-$15.9 & $-$14.9 & $-$14.7 
& $-$14.4 \\
$\lambda_3$   & $-$19.5 & $-$16.8 & $-$13.4 & $-$14.1 & $-$13.5 
& $-$11.6 \\
$\lambda_4$   & $-$21.0 & $-$19.0 & $-$17.1 & $-$16.9 & $-$16.8 
& $-$16.5 \\
$\lambda_5$   &   9.14 &   8.03 &   7.20 &   6.03 &   6.00 &   6.01 \\
$\lambda_6$   &   0.661 &  $-$3.01 &  $-$7.08 &  $-$7.32 &  $-$8.03 
& $-$10.4 \\
$\tilde{h}_A$ &   1.24 &   0.918 &   0.543 &   0.612 &   0.548 &   0.329 \\
\hline
$\chi^2$&0.26&0.38&0.77&0.92&0.94&1.2\\ \hline
$\sigma(0)$ (MeV)  & 102.1 & 75.0 & 45.0 &  84.0 & 75.0 & 45.0 \\
$\zeta_{\pi NN}$
& 1.088 &  1.056 &  1.021 &   1.050 &  1.041 &  1.012 \\
$\zeta_{\pi N\Delta}$
& 0.905 &  1.024 &  1.156 &   1.160 &  1.183 &  1.260 \\
\end{tabular}
\begin{flushleft}
$^\ast$ Constrained.
\end{flushleft}
\label{tab1}
\end{center}
\end{table}

The ratios of the effective $\pi NN$ coupling constant to the bare value,
$\zeta_{\pi NN}$, in Table \ref{tab1} show a Goldberger-Treiman discrepancy 
of a few percent. Schr\"oder et al. \cite{ph} have recently determined a 
precise result, namely 
$\zeta_{\pi NN}=1.027^{+0.012}_{-0.008}$. A similar value is obtained from 
the latest $\pi NN$ coupling constant obtained by the George Washington 
University/TRIUMF group \cite{pavan}. These values favor a $\sigma$ term 
of between
75 and 45 MeV which is in accord with several recent analyses. In Ref.
\cite{pavan} 64 MeV is quoted with an error of about 10\%. A sum rule 
determination by Olsson \cite{ols} gave 55 MeV with a 16\% error. Finally
Schr\"oder et al. \cite{ph} have indicated that the value of 45 MeV extracted 
\cite{gas} from the Karlsruhe data should be increased by 13 MeV giving
58 MeV. Lastly we note that since $\zeta_{\pi N\Delta}$ remains fairly 
close to unity the coupling is not changed too much from the bare value, 
which seems intuitively reasonable.

\begin{table}
\caption{$S$-wave scattering lengths and $P$-wave scattering volumes 
in units of $m^{-1}$ and $m^{-3}$, respectively}
\begin{center}
	\begin{tabular}{|r|rrr|rrr|r|}
	&\multicolumn{3}{c|}{$S$-matrix}&\multicolumn{3}{c|}{$K$-matrix}
&\multicolumn{1}{c|}{Experiment}\\ \cline{2-8}
	\hline
$\sigma(0)$ (MeV)	& 102.1 & 75.0& 45.0& 84.0& 75.0& 45.0& \\
	\hline
$b_0$      &  0.0051 & $-$0.0061 & $-$0.0180 &  0.0016  & $-$0.0024 
           & $-$0.0155  &  $-0.0012\pm0.0010$ \protect\cite{thom}\\
&&&&&&&$-0.0034\pm0.0007$ \protect\cite{bean}\\ \hline
$b_1$      & $-$0.0830 & $-$0.0826 & $-$0.0824 & $-$0.0847  & $-$0.0846 
           & $-$0.0843  & $-0.0895\pm0.0016$ \protect\cite{thom}\\
&&&&&&&$-0.0918\pm0.0013$ \protect\cite{bean}\\ \hline
$a_{11}$   & $-$0.0753 & $-$0.0754 & $-$0.0752 & $-$0.0769  & $-$0.0766 
           & $-$0.0757  & $-0.078\pm0.002$ \protect\cite{karl}\\
$a_{13}$   & $-$0.0290 & $-$0.0289 & $-$0.0282 & $-$0.0281  & $-$0.0280 
           & $-$0.0276  & $-0.030\pm0.002$ \protect\cite{karl}\\
$a_{31}$   & $-$0.0392 & $-$0.0398 & $-$0.0400 & $-$0.0378  & $-$0.0379 
           & $-$0.0382  & $-0.046\pm0.007$ \protect\cite{fm}\\
$a_{33}$   &  0.2180 &  0.1984 &  0.1787 &  0.1966  &  0.1919 &  0.1766  
           & $0.205\pm0.004$ \protect\cite{fm}
\label{tab2}
\end{tabular}
\end{center}
\end{table}

It is also interesting to examine the threshold results.
We give in Table \ref{tab2} the $S$-wave isoscalar and
isovector scattering lengths,  $b_0=(a_1+2a_3)/3$ and $b_1=(a_3-a_1)/3$
(with the notation $a_{2I}$) and
the $P$-wave scattering volumes ($a_{2I\,2J}$). 
The experimental values come from various sources. The $S$-wave isoscalar
scattering lengths were obtained \cite{thom,bean} from the pionic atom data 
\cite{ph} with an improved treatment of the $\pi$-$d$ scattering length. The 
$I=\thalf$ $P$-wave scattering volumes are the old Karlsruhe values 
\cite{karl}, while the $I=\thrhalf$ volumes are from a recent analysis by
Fettes and Matsinos \cite{fm}. Our calculated results for $b_0$ at the extreme 
values of $\sigma(0)$ are clearly not favored by the data which prefer
a value in the middle. Notice that, at this level of accuracy, there is a 
noticeable difference between the $S$- and $K$-matrix results. The value of 
$a_{33}$ appears to be somewhat low for the cases where $\sigma(0)=45$ MeV.
Apart from this, the remaining $P$-wave results and the values of $b_1$ 
show little sensitivity to the calculation employed and are in reasonable
agreement with the data. Of course this is not surprising since
the low energy phase shifts are included in the fit.

\subsection{IR Results}

\begin{figure}[p]
\setlength{\epsfysize}{8in}
\centerline{\epsffile{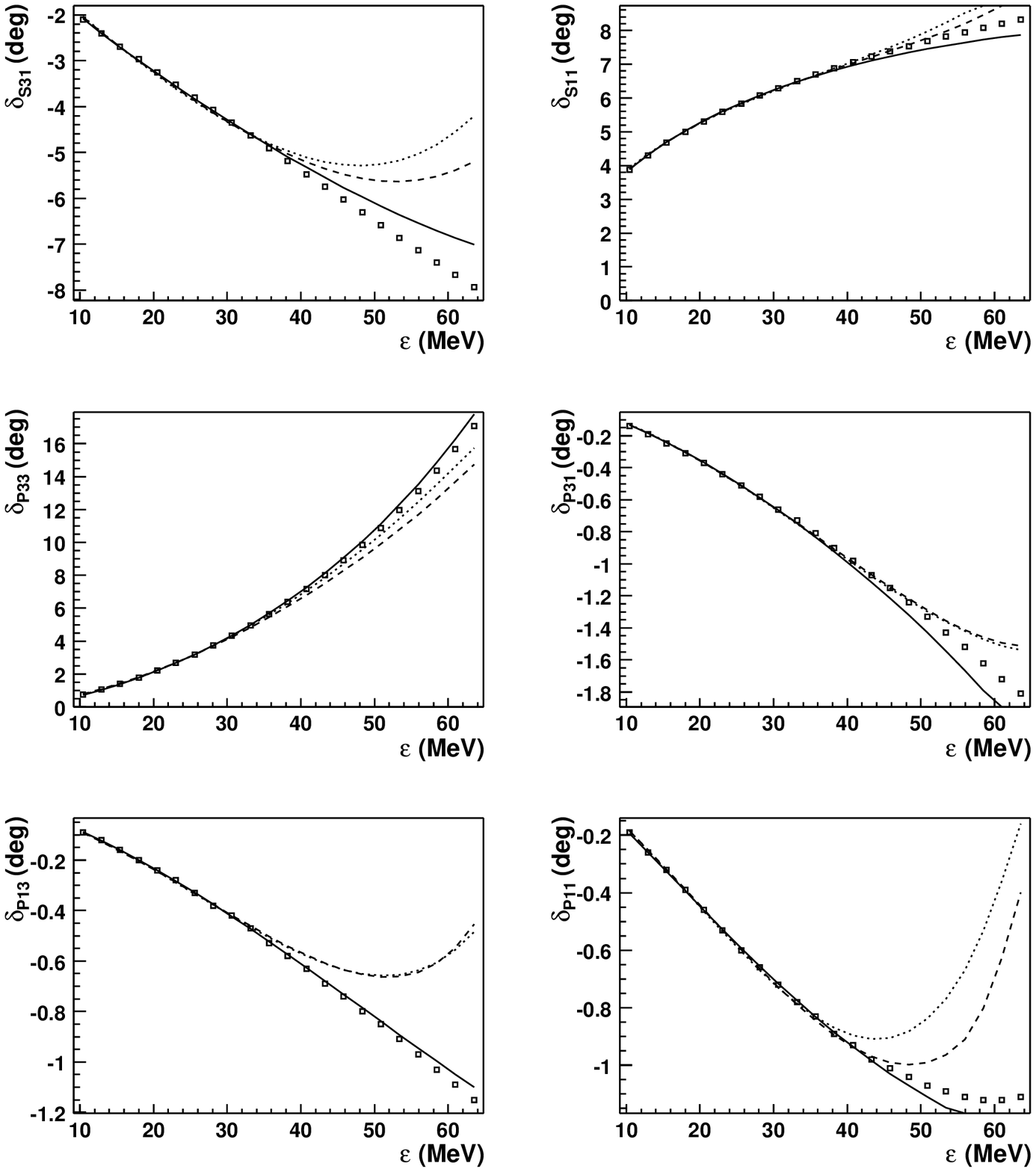}}
\caption{The $S$- and $P$-wave phase shifts from the $S$-matrix 
in the IR scheme (dashed curve) and from the $K$-matrix in the IR scheme 
(dotted) and HB scheme (solid) as a function of the pion c.m. kinetic 
energy, $\epsilon$. 
The data are from Ref. \protect\cite{ARNDT}}
\label{irsk}
\end{figure}

\begin{table}
\caption{Low energy constants, $\sigma$ term  and effective couplings 
obtained by fitting to $\epsilon=$ 40 MeV using the $S$- and $K$-matrix 
in the IR scheme and the $K$-matrix in the HB scheme}
\begin{center}
	\begin{tabular}{|r|rr|r|}
&\multicolumn{2}{c|}{IR}&HB\\
	   & $S$-matrix & $K$-matrix & $K$-matrix \\
	\hline
	$\beta_{\pi}$   &   4.23 &   5.31  &   6.12 \\
	$\kappa_{\pi}$  & $-$10.4 & $-$12.9  & $-$12.0\\
	$\kappa_1$      & $-$10.4 & $-$15.1  & $-$14.5\\
	$\kappa_2$      &  $-$0.121 &  $-$0.244  &  $-$0.287 \\
	$\lambda_1$     &  22.0 &  33.6  &  31.5\\
	$\lambda_2$     & $-$24.4 & $-$26.9  & $-$23.7 \\
	$\lambda_3$     & $-$38.2 & $-$45.3  & $-$23.9 \\
	$\lambda_4$     & $-$16.9 & $-$18.9  & $-$24.4\\
	$\lambda_5$     &  18.0 &  18.8  &   7.74 \\
	$\lambda_6$     &  19.6 &  19.1  &  $-$6.00\\
	$\tilde{h}_A$   &   0.500 &   0.664  &   1.50 \\	\hline
$\chi^2$&0.075&0.066&0.025\\ \hline
	$\sigma(0)$(MeV)& 108.9  & 119.0 &  88.4 \\
	$\zeta_{\pi NN}$&   0.790 &   0.699  &   1.043 \\
	$\zeta_{\pi N\Delta}$& 0.378 & 0.381& 1.079 \\
	\end{tabular}
\label{irint}
\end{center}
\end{table}

Turning to the IR scheme the results are disappointing. It was only found 
possible to fit the phase shift data in the low energy regime out to
$\epsilon=40$ MeV. Specifically we fitted for $\sqrt{s}$ from 1090 to 1123 
MeV in steps of 3 MeV. The IR $S$-matrix results are shown by the dashed 
curves in Fig. \ref{irsk}; here the restriction $\tilde{h}_A\leq0.5$ was 
imposed, see the discussion below. The $K$-matrix results are denoted by 
dotted curves in the figure. For contrast the solid line shows the HB results 
for a fit over the same energy range; the HB results from the $S$- and 
$K$-matrix methods are very similar and we choose to display the $K$-matrix
results. While the HB phase shifts follow the 
data approximately for $\epsilon>40$ MeV, in a number of cases the IR results 
rapidly diverge from the experimental values. The parameters corresponding 
to these results are collected in Table \ref{irint}. While the HB
parameters differ from those given in Table \ref{tab1}, particularly those 
of \ord{3}, we caution that they are not well determined by a fit over this 
limited energy range. Notice that, for the $K$-matrix cases, 
$\chi^2$ is substantially larger in the IR scheme than the HB one.
In the IR case $\zeta_{\pi NN}$ differs substantially 
from unity, implying a huge Goldberger-Treiman discrepancy. This is 
unphysical and is a signal of the contortions that the parameters are going 
through in order to fit over even this limited range of energies. The 
quantity $\zeta_{\pi N\Delta}$ is also much less than unity implying an 
effective $\pi N\Delta$ coupling very different from the bare value.

\begin{table}
\caption{Comparison of IR results for the real parts of various integrals and 
$\Pi_\Delta$ in the $s$-channel with the corresponding HB approximation.}
\begin{center}
	\begin{tabular}{|l|rr|rr|}
	& \multicolumn{2}{c|}{$\sqrt{s}=1200$ MeV} &
		\multicolumn{2}{c|}{$\sqrt{s}=1100$ MeV} \\ 
Integral & IR & HB & IR & HB\\
	\hline
	$\bar{{\cal I}}_{11}(s,M_{\Delta})$  
		&   0.338 &   0.357 &   0.148 &   0.120  \\
	$\bar{{\cal I}}_{11}^{(2)}(s,M_{\Delta})/\bar{M}^2$  
		&  0.00193 &  0.00199 &  0.000314 & 0.000192  \\
	$\bar{{\cal I}}_{12}^{(3)}(s,M_{\Delta})$  
		& $-$0.00127 & $-$0.00413 &  $-$0.00797 & $-$0.0103  \\
	$\bar{M}^2\bar{{\cal I}}_{13}^{(3)}(s,0,M_{\Delta},M,M_{\Delta})$  
		&  0.0798 &   0.0453 &   0.1020 &  0.0568  \\
	$\Pi_{\Delta}(s)/\bar{M}^2$ for $\tilde{h}_A^2=0.1$
		&  0.000287 &   0.00546 &  0.0230 &  0.0696  
\label{irsint}
\end{tabular}
\end{center}
\end{table}

\begin{table}
\caption{Comparison of IR results for the real parts of various integrals 
and $\Pi_\Delta$ in the $u$-channel, for $u=u_{\rm min}$, with the 
corresponding HB approximation.}
\begin{center}
	\begin{tabular}{|l|rr|rr|}
	& \multicolumn{2}{c|}{$\sqrt{s}=1200$ MeV} &
\multicolumn{2}{c|}{$\sqrt{s}=1100$ MeV} \\ 
	& \multicolumn{2}{c|}{$\sqrt{u_{\text{min}}}=$601.7 MeV} &
\multicolumn{2}{c|}{$\sqrt{u_{\text{min}}}=$769.5 MeV} \\ 
Integral		& IR & HB & IR & HB\\
	\hline
	$\bar{{\cal I}}_{11}(u,M_{\Delta})$  
		&   $-$6.51 &   $-$1.42 &   $-$2.39 &   $-$0.934  \\
	$\bar{{\cal I}}_{11}^{(2)}(u,M_{\Delta})/\bar{M}^2$  
		&    1.34 &    0.0826 &  0.173 &  0.0316  \\
	$\bar{{\cal I}}_{12}^{(3)}(u,M_{\Delta})$  
		&    0.840 &  0.110 &  0.180 &   0.0500\\
	$\bar{M}^2\bar{{\cal I}}_{13}^{(3)}(u,0,M_{\Delta},M,M_{\Delta})$  
		&    0.597 &   0.167 &   0.321 &  0.130  \\
	$\Pi_{\Delta}(u)/\bar{M}^2$ for $\tilde{h}_A^2=0.1$
		&    $-$0.751 &   0.319 &  0.0213 &  0.262  
\end{tabular}
\label{iruint}
\end{center}
\end{table}

We have, of course, carried out a number of additional IR calculations 
with a view to improving the results. For example constraining the 
value of the $\sigma$ term to be 75 or 45 MeV increases the value of 
$\chi^2$, but the results are qualitatively very similar to those 
discussed above. In order to see whether the problem was due to the 
$\Delta$ resonance contributions we removed them entirely by setting 
the coupling $h_A$ to zero. As expected the parameters changed 
substantially, but the fit to the phase shifts remained qualitatively 
the same and in fact $\chi^2$ increased somewhat. Thus we have not been
able to qualitatively improve the IR results discussed in detail above.

In order to further investigate the reason for the problem with the 
IR approach we 
compare the IR polarization $\Pi_\Delta$ of Eq. (\ref{jkp}) and a few IR 
integrals (defined in the Appendix) with their HB counterparts. In Table
\ref{irsint} we make the comparison in the $s$-channel. For the most part 
the IR and HB results are quite comparable. The picture changes for the
$u$-channel in Table \ref{iruint}, where for a given $s$ we have chosen
the minimum value, $u_{\rm min}=2(M^2+m^2)-s$, so as to enhance the 
contrast.
Here the IR integrals are much larger than the corresponding HB results,
by a factor of 6 on average. Further $\Pi_\Delta(u)$ changes sign. Since
$\Delta_R(u)=1/[u-M_\Delta^2-\Pi_\Delta(u)]$ and $u-M_\Delta^2$ is 
negative this change of sign can produce a pole; this was the reason for our 
restriction of $\tilde{h}_A\leq0.5$ in the $S$-matrix IR fit. Of course the 
problem does not arise if $\Pi_\Delta(u)$ is expanded out of the 
denominator as in the $K$-matrix approach, although the mathematical 
justification for doing so is slight. The problem does not arise in the HB 
case where $\Pi_\Delta(u)$ is always positive.

\begin{figure}
\setlength{\epsfysize}{3.5in}
\centerline{\epsffile{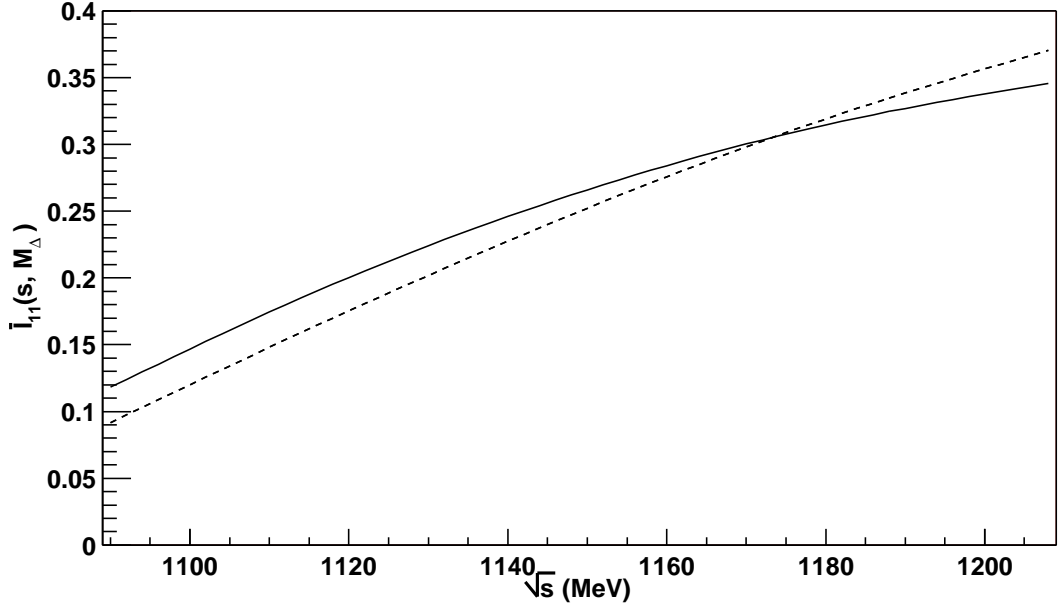}}
\caption{Comparison of the integral 
$\bar{{\cal I}}_{11}(s,M_\Delta)$ evaluated exactly (solid curve) 
with the HB approximation (dashed curve) as a function of $\sqrt{s}$.}
\label{sint}
\end{figure}

\begin{figure}
\setlength{\epsfysize}{3.5in}
\centerline{\epsffile{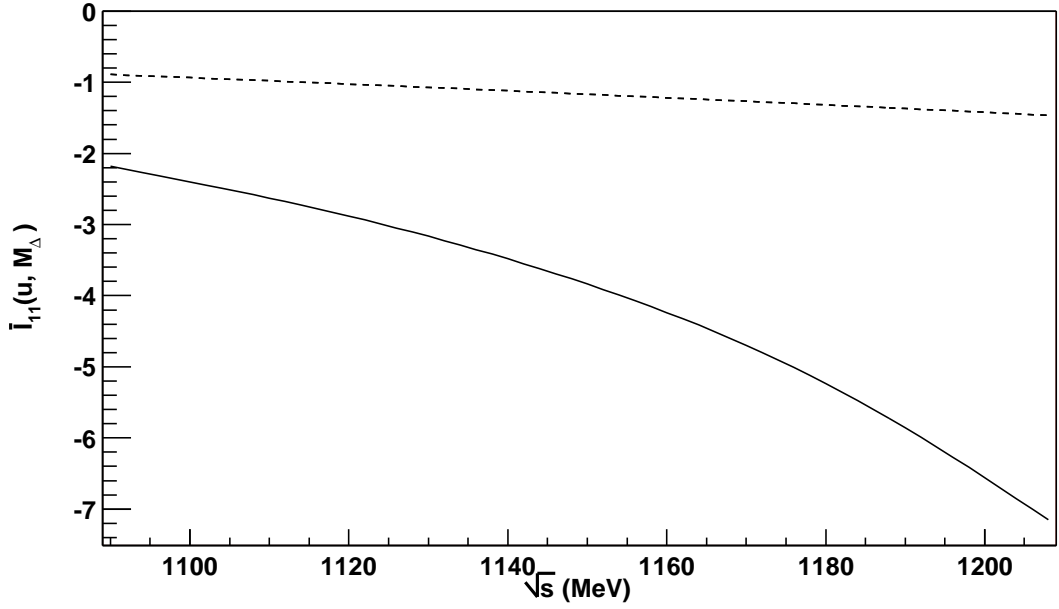}}
\caption{Comparison of the integral 
$\bar{{\cal I}}_{11}(u,M_\Delta)$, for $u=u_{\rm min}$, evaluated 
exactly (solid curve) with the HB approximation (dashed curve)
as a function of $\sqrt{s}$.}
\label{uint}
\end{figure}

It is also useful to make the point graphically by studying the basic 
integral involving one baryon and one meson propagator. Figure \ref{sint}
shows $\bar{\cal I}_{11}(s,M_\Delta)$ as a function of $\sqrt{s}$. It is 
observed that the IR integral (solid line) and the HB approximation thereto 
(dashed curve) agree quite closely. Figure \ref{uint} shows the 
corresponding result in the $u$-channel; 
$\bar{\cal I}_{11}(u_{\rm min},M_\Delta)$ as a function of 
$\sqrt{s}$. It is seen that the magnitude of the IR integral (solid line)
is much larger than that of its HB counterpart (dashed curve); note the 
ordinate scale here in comparison to Fig. \ref{sint}. The IR integral is 
governed by the value of $x$ (see Appendix) which in the $s$-channel is 
$(s-M_\Delta^2+m^2)/(2m\sqrt{s})$. In the HB approximation this is 
replaced by $(s-M_\Delta^2)/(2m\bar{M})=(\eta-\delta)/m$, in the notation
of I. At $\sqrt{s}=1200$ MeV, $x$ is 
$-0.18$ ($-0.26$) in the IR (HB) case. Correspondingly in the $u$-channel, 
at $\sqrt{u_{\rm min}}=601.7$ MeV, $x$ is $-6.79$ ($-3.83$) in the IR (HB) 
case and this large difference is the heart of the problem with the IR 
approach. However, it is disquieting that the rather reasonable HB results 
are achieved by the replacement of $\sqrt{u}$ in the denominator by 
the square root of average baryon mass, $\sqrt{\bar{M}}$,
since this can be quite a rough approximation.

\section{Conclusions}

We have performed a least-squares fit to the $S$- and $P$-wave phase 
shift data using chiral perturbation theory to \ord{3} with explicit
$\pi$, $N$ and $\Delta$ fields. The heavy baryon formulation used here 
represents an improvement compared to I in that
the polynomial terms were included explicitly and the renormalization 
scale dependance was subsumed into the low energy constants. We contrasted
an ``$S$-matrix" method (as in I) with a $K$-matrix analysis (as in Ref. 
\cite{fm3nd}). The former seems to be more physical since it
employs dressed $\Delta$ propagators which are finite, while the latter
uses bare propagators which give a divergent scattering amplitude at the 
$\Delta$ resonance energy. Broadly speaking the two methods give similar 
results and allow a fit to the phase shift up to a c.m. pion kinetic 
energy, $\epsilon$, of 100 MeV. In detail, however, there are some 
differences. For instance, the value of the $\sigma$ term (if 
unconstrained) and the Goldberger-Treiman discrepancy are smaller with the 
$K$-matrix.

Our main interest was to see whether the infrared regularization scheme, 
which is manifestly Lorentz invariant, could provide an equally satisfactory
account of the phase shift data. It could not. We were only able to fit the 
data up to $\epsilon=40$ MeV and the resulting Goldberger-Treiman 
discrepancy was ridiculously large. The difficulty was not due to the 
explicit inclusion of the $\Delta$ resonance, but rather stemmed from the 
$u$ channel where the magnitudes of the integrals were much larger than the 
corresponding heavy baryon ones. This leads to uneasiness regarding the 
latter since they are obtained by approximating the former. The heavy 
baryon scheme thus pushes significant contributions into higher orders where 
they are combined with the other terms of that order. The only information 
we have as to whether the net effect is small comes from the comparison of
third and fourth orders carried out by Fettes and Mei\ss ner \cite{fm4}. 
In some cases there are sizeable differences, although these authors 
suggest that in most cases the chiral series appears convergent.
Overall, in our opinion, the description of pion-nucleon scattering via
chiral perturbation theory is not yet in a satisfactory state.

\section*{Acknowledgements}
We acknowledge support from the Department of Energy under grant 
No. DE-FG02-87ER40328.

\appendix
\section*{}

Integrals involving only meson propagators need no 
discussion since the standard results are used and integrals involving
only baryon propagators give no contribution in the IR or HB schemes.
Infrared regularization is needed for integrals involving both baryon 
and meson propagators. Here we give only those integrals that are referred 
to in the main text, together with their HB reduction. The integrals needed
can be derived from the basic infrared integral which has been discussed by 
Becher and Leutwyler \cite{bl} (note that our conventions differ 
from theirs), see also I. Explicitly this is
\begin{eqnarray}
&&i\mu^{4-d}\int_I\frac{d^d\ell}{(2\pi)^d}\frac{1}{[(\ell+k)^2-{\cal M}^2
+i\epsilon]
[\ell^2-m^2+i\epsilon]}\nonumber\\
&&=i\mu^{4-d}\int\frac{d^d\ell}{(2\pi)^d}\frac{1}{
[2k\cdot\ell+k^2-{\cal M}^2+m^2+i\epsilon][\ell^2-m^2+i\epsilon]}
\nonumber\\
&&={\cal I}_{11}^L(s,{\cal M})+(4\pi)^{-2}
\bar{\cal I}_{11}(s,{\cal M})\;,
\end{eqnarray}
where $s=k^2$. The total integral, without the subscript $I$, will have, in 
addition to the infrared part, a regular part but this can be discarded 
since it can be expanded with the various terms absorbed into the low 
energy constants. The divergent IR contribution is
\begin{eqnarray}
&&{\cal I}_{11}^L(s,{\cal M})=\frac{(s-{\cal M}^2+m^2)}{2s}
\left[2L+\frac{1}{(4\pi)^2}\ln\frac{m^2}{\mu^2}\right]\;,\nonumber\\
&&{\rm where}\quad L=\frac{1}{32\pi^2}\left(\frac{2}
{d-4}+\gamma-1-\ln4\pi\right)\;,
\end{eqnarray}
with $\gamma$ denoting Euler's constant. The real part of the finite piece 
is
\begin{eqnarray}
\Re\bar{\cal I}_{11}(s,{\cal M})&=&\left\{\matrix{\frac{m}{\sqrt{s}}
\left[-x+2\sqrt{1-x^2}\cos^{-1}(-x)\right]&\qquad(|x|\leq1)\cr
\frac{m}{\sqrt{s}}\left[-x+\sqrt{x^2-1}\ln\frac{x+\sqrt{x^2-1}}
{x-\sqrt{x^2-1}}\right]&\qquad(|x|\geq1)\cr}\right. \nonumber\\
&\rightarrow&-\frac{1}{\bar{M}}J\left(\frac{s-{\cal M}^2}{2\bar{M}}
\right)\;,
\end{eqnarray}
where $x=(s-{\cal M}^2+m^2)/(2m\sqrt{s})$. The last line here makes contact 
with the HB integral $J$ in I for which $\sqrt{s}$ in the denominator is
approximated in leading order by the average baryon mass 
$\bar{M}=\thalf(M+M_\Delta)$. The imaginary part of the IR integral is
\begin{equation}
\Im\bar{\cal I}_{11}(s,{\cal M})= 
-\frac{\pi}{s}\sqrt{(s-{\cal M}^2+m^2)^2-4sm^2}\:\theta(s-[{\cal M}+m]^2)\;,
\end{equation}
where $\theta$ denotes the Heaviside step function.

In the following we shall implictly give only the real part of the IR 
integral and the $i\epsilon$ in the integrand denominators will be 
suppressed. For the integral 
\begin{equation}
i\mu^{4-d}\int_I\frac{d^d\ell}{(2\pi)^d}\frac{\ell_\mu}
{[(\ell+k)^2-{\cal M}^2][\ell^2-m^2]}
={\cal I}_{11}^{(1)}(s,{\cal M})k_\mu\;.
\end{equation}
The finite part is $(4\pi)^{-2}\bar{\cal I}_{11}^{(1)}$,
with
\begin{equation}
\bar{\cal I}_{11}^{(1)}(s,{\cal 
M})=-\left[\frac{s-{\cal M}^2+m^2}{2s}\right]
\bar{\cal I}_{11}(s,{\cal M})
\rightarrow\left[\frac{s-{\cal M}^2}{2\bar{M}^3}
\right]J\left(\frac{s-{\cal M}^2}{2\bar{M}}\right)\;,
\end{equation}
with the last expression giving the HB reduction.

For the integral 
\begin{equation}
i\mu^{4-d}\int_I\frac{d^d\ell}{(2\pi)^d}\frac{\ell_\mu\ell_\nu}
{[(\ell+k)^2-{\cal M}^2][\ell^2-m^2]}
={\cal I}_{11}^{(2)}(s,{\cal M})g_{\mu\nu}
+{\cal I}_{11}^{(3)}(s,{\cal M})k_\mu k_\nu\;.
\end{equation}
The finite part of the first term is $(4\pi)^{-2}\bar{\cal I}_{11}^{(2)}$,
with
\begin{eqnarray}
\bar{\cal I}_{11}^{(2)}(s,{\cal 
M})&=&\left[\frac{4m^2s-(s-{\cal M}^2+m^2)^2}{12s}\right]
\bar{\cal I}_{11}(s,{\cal M})
-\frac{(s-{\cal M}^2)}{36s^2}\left[6m^2s-(s-{\cal M}^2)^2
\right]\nonumber\\
&\rightarrow&-\frac{1}{3\bar{M}}\left[m^2\!-\frac{(s-{\cal M}^2)^2}
{4\bar{M}^2}\right]\!J\!\left(\frac{s-{\cal M}^2}{2\bar{M}}\right)
-\frac{(s-{\cal M}^2)}{18\bar{M}^2}\left[3m^2\!-\frac{(s-{\cal 
M}^2)^2}{2\bar{M}^2}\right],
\end{eqnarray}
with the last line giving the HB reduction.

One of the integrals involving one pion and two baryon denominators is
\begin{eqnarray}
&&i\mu^{4-d}\int_I\frac{d^d\ell}{(2\pi)^d}\frac{\ell_\mu\ell_\nu}
{[(\ell+p'+q')^2-{\cal M}_A^2][(\ell+p')^2-{\cal M}_B^2][\ell^2-m^2]}
\nonumber\\
&&={\cal I}_{12}^{(3)}(s,{\cal M}_A,{\cal M}_B)g_{\mu\nu}+\ldots\;.
\end{eqnarray}
The finite part is $(4\pi)^{-2}\bar{\cal I}_{12}^{(3)}$, with
\begin{eqnarray}
&&\bar{{\cal I}}_{12}^{(3)}(s,{\cal M}_A,{\cal M}_B)=
\left(\frac{\partial}{\partial{\cal M}_A^2}
+\frac{\partial}{\partial{\cal M}_B^2}\right)\int\limits_0^1 dz\,
\bar{{\cal I}}_{11}^{(2)}(\bar{k}^2,\bar{\cal M})\;,\nonumber\\
&&{\rm where}\quad\bar{k}^2=m^2z^2+(s-M^2-m^2)z+M^2\;,\nonumber\\
&&{\rm and}\quad\bar{\cal M}^2=m^2z^2+({\cal M}_A^2-{\cal M}_B^2-m^2)z
+{\cal M}_B^2\;. \label{ir123}
\end{eqnarray}
This was evaluated by performing the differentiation analytically and 
carrying out the integration numerically using Gaussian quadrature.
It is reduced to HB form by dropping terms involving $m^2$ from
the expression $\quad\bar{k}^2-\bar{\cal M}^2+m^2$ and setting
$\quad\bar{k}^2\simeq M^2$ in denominators. This allows the integral to be 
performed, giving
\begin{eqnarray}
&&\bar{{\cal I}}_{12}^{(3)}(s,M_\Delta,M)\rightarrow
-\frac{1}{6\bar{M}^2}\left[F_3(\eta-\delta,0)-m^2
+\ttwoth(\eta-\delta)^2\right]\;,\nonumber\\
&&{\rm where}\quad F_3(\alpha,\beta)=\frac{1}{\alpha-\beta}
\left[(\alpha^2-m^2)J(\alpha)-(\beta^2-m^2)J(\beta)\right]\;,
\end{eqnarray}
and we recall the definitions of I: $2\bar{M}\eta=s-M^2$ and
$\delta=M_\Delta-M$.

The integral involving one pion and three baryon denominators is
\begin{eqnarray}
&&i\mu^{4-d}\int_I\frac{d^d\ell}{(2\pi)^d}\frac{\ell_\mu\ell_\nu}
{[(\ell+p')^2-{\cal M}_A^2][(\ell+p'+q')^2-{\cal M}_B^2]
[(\ell+p)^2-{\cal M}_C^2][\ell^2-m^2]}
\nonumber\\
&&={\cal I}_{13}^{(3)}(s,t,{\cal M}_A,{\cal M}_B,{\cal M}_C)g_{\mu\nu}
+\ldots\;.
\end{eqnarray}
The finite part is $(4\pi)^{-2}\bar{\cal I}_{13}^{(3)}$, with
\begin{eqnarray}
&&\bar{{\cal I}}_{13}^{(3)}(s,t,{\cal M}_A,{\cal M}_B,{\cal M}_C)=
\left(\frac{\partial}{\partial{\cal M}_A^2}+
\frac{\partial}{\partial{\cal M}_B^2}+\frac{\partial}{\partial{\cal 
M}_C^2}\right)^{\!2}\int\limits_0^1dx x
\int\limits_0^1dy\,\bar{\cal I}_{11}^{(2)}(\bar{k}^2,\bar{\cal M})
\nonumber\\
&&{\rm where}\quad\bar{k}^2=\left[s-M^2-m^2+m^2x(1-y)\right]x(1-y)
-txy(1-x)+M^2\;,\nonumber\\
&&{\rm and}\quad\bar{\cal M}^2=m^2x^2(1-y)^2+\left[{\cal M}_A^2
-{\cal M}_B^2+m^2-t(1-x)\right]xy\nonumber\\
&&\qquad\qquad\qquad+({\cal M}_B^2-{\cal M}_C^2-m^2)x
+{\cal M}_C^2\;.
\end{eqnarray}
This was evaluated with a combination of numerical differentiation and 
Gaussian numerical integration.
The HB reduction is carried out by making similar approximations to those
mentioned above and this allows both integrals to be performed. For the 
cases
mentioned in the text this gives 
\begin{eqnarray}
&&\bar{{\cal I}}_{13}^{(3)}(s,t,M_\Delta,M,M_\Delta)\rightarrow
\frac{1}{12\bar{M}^3}\left[F_1(\eta,-\delta)+\ttwoth(\eta-2\delta)\right]
\;,\nonumber\\
&&{\rm where}\quad F_1(\alpha,\beta)=\frac{
(\alpha^2-m^2)J(\alpha)+(2\beta^2-3\alpha\beta+m^2)J(\beta)
+(\alpha-\beta)(2\beta^2-m^2)}{(\alpha-\beta)^2}\,,
\end{eqnarray}
and
\begin{eqnarray}
&&\bar{{\cal I}}_{13}^{(3)}(s,t,M_\Delta,M,M)\rightarrow
\frac{1}{12\bar{M}^3}\left[F_2(\eta,-\delta)+\ttwoth(\eta-\delta)\right]
\;,\nonumber\\
&&{\rm where}\quad F_2(\alpha,\beta)=\frac{
(\alpha^2-m^2)\beta J(\alpha)-(\beta^2-m^2)\alpha J(\beta)
+\pi m^3(\alpha-\beta)}{\alpha\beta(\alpha-\beta)}\,.
\end{eqnarray}

\end{document}